\documentclass[preprint,number]{elsarticle}
\usepackage[]{graphicx}
\usepackage{amsmath}
\usepackage{amsfonts}
\usepackage{amssymb}
\usepackage{multirow}
\def\Journal#1#2#3#4{{#1} {\bf #2}, #4 (#3)}

\def\NIMA{Nucl.\,Instrum.\,Methods A}

\def\NPB{Nucl.\,Phys. B}

\def\PRL{Phys.\,Rev.\,Lett.}
\def\PRD{Phys.\,Rev. D}
\def\PRC{Phys.\,Rev. C}

\def\JLT{J.\,Low\,Temp.\,Phys.}

\def\Re{$^{187}$Re}

\def\agre{AgReO$_4$}
\def\Tr{$^3$H}
\def\mn{$m_\nu$}
\def\mnsq{$m_\nu^2$}

\def\mus{$\mu$s}

\def\de{$\Delta E$}
\def\fwhm{$_{\mathrm{FWHM}}$}
\begin{document}
\begin{frontmatter}
\title{Expectations for a new calorimetric\\neutrino mass experiment}
\author[unimib,mib]{A. Nucciotti\corref{cor1}}
\ead{angelo.nucciotti@mib.infn.it}
\author[unimib,mib]{E. Ferri} 
\author[mib]{O. Cremonesi}
\address[unimib]{Dipartimento di Fisica dell'Universit\`a di Milano-Bicocca, Milano, Italia}
\address[mib]{INFN, Sezione di Milano-Bicocca, Milano, Italia}
\cortext[cor1]{Corresponding author}
\begin{abstract}
A large calorimetric neutrino mass experiment using thermal detectors is expected to play a crucial role in the challenge for directly assessing the neutrino mass. 
We discuss and compare here two approaches to the estimation of the experimental sensitivity of such an experiment. 
The first method uses an analytic formulation and allows to readily obtain a sensible estimate over a wide range of experimental configurations. The second method is based on a frequentist Montecarlo technique and is more precise and reliable. 
The Montecarlo approach is then exploited to study the main sources of systematic uncertainties peculiar to calorimetric experiments.
Finally, the tools are applied to investigate the optimal experimental configuration for a calorimetric experiment with Rhenium based thermal detectors.
\end{abstract}
\end{frontmatter}
\section{Introduction}
Assessing the neutrino mass scale is one of the major challenges in today particle physics and astrophysics. This 
requires to measure the mass of one of the three neutrinos and the  kinematical neutrino mass
measurement is the only model independent method.
In particular, the electron anti-neutrino 
mass can be measured by precisely analyzing the kinematics of electrons emitted in beta decays. 
In practice this means measuring the minimum energy carried away by the anti-neutrino, i.e. its rest mass, by observing the highest energy electrons emitted in the decay.
To date, the study of the \Tr\ beta decay end-point by means of electrostatic spectrometers has proved to be the
most  sensitive approach, yielding an upper limit on the electron anti-neutrino mass of
2.2\,eV\,\cite{MainzTroitsk}. Starting from 2012 the new experiment KATRIN will  analyze the \Tr\ beta decay end-point  with a
much more sensitive electrostatic spectrometer and with an expected  statistical sensitivity of about 0.2\,eV\,\cite{KATRIN}

However, these spectrometric experiments suffer from many systematic uncertainties because the measured electron energy has to be corrected for the
energy lost in exciting atomic and molecular states, 
in crossing the source, in scattering through the spectrometer, and more. 
To avoid these uncertainties it was proposed to embed the beta source in a detector and to perform a so called calorimetric measurement.
Ideally, in such a configuration for each decay the detector measures all the energy released except for the energy carried away by the neutrino.\footnote{In practice particles emitted at the detector surface or with enough energy may also escape detection. In most cases, however, only a small fraction of the decays are affected by such an effect.} 

A drawback of calorimetry is that a calorimeter is forced to detect all the beta decays while only the ones very close to the end-point $E_0$ are useful for measuring the neutrino mass. The fraction of useful decays in a small interval \de\ below $E_0$ is approximately given by $(\Delta E/E_0)^3$, 
therefore it pays out to select a beta decaying isotope with the lowest $E_0$ value. 
In the past, calorimetric neutrino mass experiments have been performed implanting \Tr\ in Silicon diode detectors \cite{Simpson}. The use of \Re\ as beta source seems more promising since it is the beta-active nuclide with the second lowest known transition energy ($E_0\sim2.5$\,keV). 
In the '80s  S.\,Vitale proposed to realize calorimetric neutrino mass using \Re\ as beta source and exploiting the thermal
detection technique \cite{Vitale}. Since then two experiments of this kind has been carried out: the MANU\cite{MANU-PRC,MANU} and MIBETA\cite{MIBETA-PRL,MIBETA} experiments. MANU used one detector with a NTD thermistor glued to a 1.6\,mg metallic rhenium single crystal, while MIBETA used an array of ten silicon implanted thermistors with \agre\ crystals for a total mass of about 2.2\,mg. The two experiments collected statistics corresponding to 10$^7$ and 1.7$\times 10^7$ decays respectively, yielding
limits on \mn\ of about 26\,eV at 95\% CL and 15\,eV at 90\% CL respectively.

Recent developments in the thermal detection technique let think about a new very large calorimetric experiment aiming at a sub-eV sensitivity: this it what the MARE project is about \cite{MARE, MARE-ltd12}.

In this paper we present a comprehensive discussion of the potential sensitivity to the neutrino mass for a calorimetric experiment. 
First, through an analytical approach, we derive an algorithm to assess the statistical sensitivity for a given experimental configuration.
Then a Montecarlo method is described which  allows to get more precise statistical sensitivity estimates. The results of the analytic approach are then validated through the comparison with the Montecarlo results over a wide range of experimental parameters.
The second part of this paper focuses on the systematic uncertainties  peculiar to the calorimetric technique by applying extensively the Montecarlo approach to their investigation. 
We conclude with a discussion of the possible experimental configurations for future large scale calorimetric experiments.
\section{Statistical sensitivity}
\subsection{The analytic approach}
\label{sec:analytic}
\begin{figure}[ht!]
\begin{center}
\includegraphics[width=0.8\linewidth, clip]{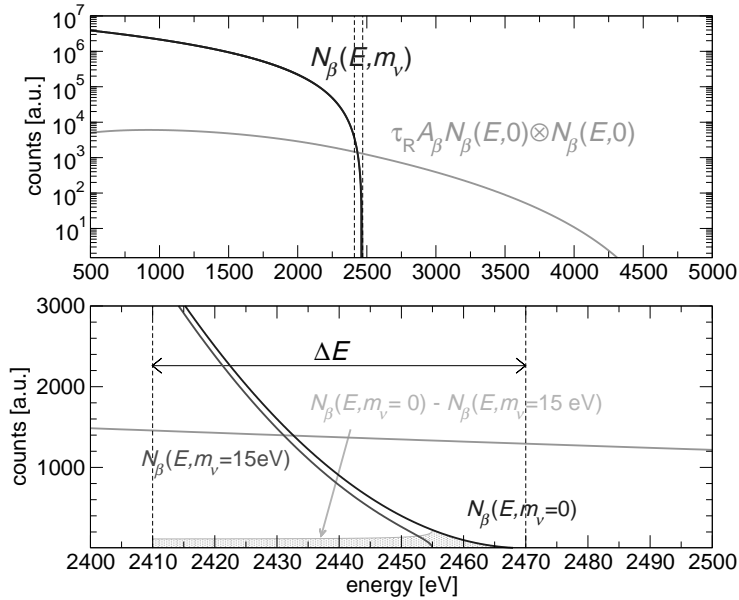}%
\caption{\label{fig:sensi} Higher panel: beta spectrum as in (\ref{eq:quad}) compared with pile-up spectrum (\ref{eq:npp}). Lower panel: zoom around the end point, with a comparison between 0 and finite-neutrino-mass beta spectra}
\end{center}
\end{figure}

In the following we derive an approximate analytic expression for the statistical sensitivity of a calorimetric neutrino mass experiment.
The primary effect of a finite mass $m_\nu$ on the beta spectrum is to cause the
spectrum to turn more sharply down to zero a distance $m_\nu$ below the end-point
$E_0$ (lower panel of Figure\,\ref{fig:sensi}). 
To rule out a finite mass, we must be sensitive to the number of counts expected in this interval. 
The fraction of the total spectrum within an interval $\Delta E$ below the end-point $E_0$ is given by
\begin{equation}\label{eq:frac}
F_{\Delta \! E}(m_\nu) =\int^{E_0}_{E_0-\Delta \! E} N_\beta(E,m_{\nu}) d \! E 
\end{equation}
where $N_\beta(E,m_{\nu})$ is the beta energy spectrum  for a
neutrino mass $m_{\nu}$ and normalized to unity. The signal to detect in counts is therefore
\begin{equation}\label{eq:sig}
signal =A_\beta N_{det} |F_{\Delta \! E}(m_\nu)-F_{\Delta \! E}(0)|t_M
\end{equation}
where $A_\beta$ is the single detector source activity, $N_{det}$ is the number of identical detectors 
and $t_M$ is the measuring time (see lower panel of Figure\,\ref{fig:sensi}).

The detection of this signal is impaired by the {\it noise} caused by the statistical fluctuations of the total measured spectrum
in the  interval $\Delta E$. For a calorimeter the total measured spectrum is obtained summing up -- from all detectors -- the beta decay events, the counts due to unresolved pile-up of two or more decays, and any additional background counts. 

As a first approximation we can neglect the pile-up of more than two decays.
We can then crudely approximate the pile-up spectrum by
assuming a constant pulse-pair resolving time, $\tau_R$, such that events with greater separation are
always detected as being doubles, while those at smaller separations are always interpreted as
singles with an apparent energy equal to the sum of the two events. In fact, the resolving time
will depend on the amplitude of both events, and the sum amplitude will depend on the
separation time and the filter used, so a proper calculation would have to be done as a Monte
Carlo with the actual filters and pulse-pair detection algorithm being used. However, this
approximation is good enough to get the correct scaling and an approximate answer.

The parameter $\tau_R$ is related to the detector signal bandwith and high frequency signal-to-noise ratio: in practice 
$\tau_R$ is of the order of the detector rise time.

The two event pile spectrum is given by
\begin{equation}\label{eq:npp}
N_{pp}(E)=(1-e^{-A_\beta \tau_R}) N_\beta(E,0) \otimes N_\beta(E,0)
\end{equation}\noindent 
and the fraction of this spectrum within the interval $\Delta E$ below the end-point $E_0$ is obtained by
\begin{equation}\label{eq:frac_pp}
F^{pp}_{\Delta E} = \int^{E_0}_{E_0-\Delta \! E}N_{pp}(E) d \! E \approx \tau_R A_\beta \int^{E_0}_{E_0-\Delta \! E} N_\beta(E,0) \otimes N_\beta(E,0)d \! E 
\end{equation}\noindent  
where, at first order, $\tau_R A_\beta$ is the probability for the two event pile-up to occur, i.e.
the fraction of unresolved pile-up events, $f_{pp}$.
From (\ref{eq:frac}) and (\ref{eq:frac_pp}) one can write the {\it noise} in counts as
\begin{equation}\label{eq:noise}
noise = {\sqrt{A_\beta N_{det}(F_{\Delta \! E}(0)+F^{pp}_{\Delta E})  t_M   +  N_{det} b \Delta E t_M } }
\end{equation}\noindent  
where $b$ is the average background counting rate for unit energy  and for a single detector.
We can then write the signal to noise ratio in a region
within  $\Delta E$ of the end-point $E_0$ as
\begin{equation}\label{eq:S2N}
\frac{signal}{noise} = \sqrt{ A_\beta N_{det}t_M} \frac{|F_{\Delta \! E}(m_\nu)-F_{\Delta \! E}(0)|}{\sqrt{F_{\Delta \! E}(0)+F^{pp}_{\Delta E}+b\Delta \! E/A_\beta}}
\end{equation}\noindent
It is now useful to introduce the exposure $T=N_{det}t_M$ and total number of events or total statistics of the experiment $N_{ev}=A_\beta N_{det} t_M$. \\
The value of $m_\nu$ which makes this ratio equal to 1.7 is the sensitivity at 90\% confidence level, $\Sigma_{90}(m_\nu)$. Therefore
one has to solve for $m_\nu$ the following equation 
\begin{equation}\label{eq:S2Neq}
\sqrt{ N_{ev}} \frac{|F_{\Delta \! E}(m_\nu)-F_{\Delta \! E}(0)|}{\sqrt{F_{\Delta \! E}(0)+F^{pp}_{\Delta E}+ b\Delta \! E/A_\beta}}  = 1.7
\end{equation}\noindent

To evaluate (\ref{eq:S2Neq}), we can consider approximate expressions for $F_{\Delta \! E}(m_\nu)$ and $F^{pp}_{\Delta E}$.
In particular if we restrict ourselves to \Re, which has a first forbidden unique beta transition,  we can make use of the following empirical spectrum
\begin{equation}\label{eq:quad}
N_\beta(E,m_{\nu}) \approx \frac{3}{E_0^3} (E_0-E)^2 \sqrt{1- \frac{m^2_\nu}{(E_0-E)^2}}
\end{equation}
which is an extremely good approximation of the expected theoretical shape  
\cite{simkovic} as well as a perfect description of the experimental observations \cite{MIBETA,MANU}.

For a null mass, from (\ref{eq:quad}) we can derive 
\begin{equation}\label{eq:F0}
F_{\Delta \! E}(0) = \left( \frac {\Delta \! E} {E_0} \right)^3
\end{equation}\noindent 
while, for a small but finite mass $m_\nu$, using a second order expansion in $m_{\nu}/\Delta E$ we have approximately
\begin{equation}\label{eq:Fm}
F_{\Delta \! E}(m_\nu) \approx F_{\Delta \! E}(0) \left(1 - \frac {3 m_{\nu}^2} {2 \Delta \! E^2} + \frac{3m_{\nu}^4 }{8 \Delta \! E^4} \right)
\end{equation}\noindent 
For the pile-up spectrum (\ref{eq:npp}) using (\ref{eq:quad}) we can calculate (between 0 and $E_0$)
\begin{equation}\label{eq:npp_re}
N_{pp}(E)= (1-e^{-A_\beta \tau_R}) \frac{1}{E_0}
\left(9\frac{E}{E_0}-18\frac{E^2}{E_0^2}+12\frac{E^3}{E_0^3}-3\frac{E^4}{E_0^4}+\frac{3}{10}\frac{E^5}{E_0^5} \right) 
\end{equation}
Between $E_0$ and $2E_0$ the expression for $N_{pp}(E)$ is more complicated and it is of no
use in this context.
Substituting (\ref{eq:npp_re}) in (\ref{eq:frac_pp}) and carrying out the integration, we obtain
\begin{equation}\label{eq:frac_pp_re}
F^{pp}_{\Delta E}  =  f_{pp} \frac{1}{20}
\left(6\frac{\Delta E}{E_0}+ 15\frac{\Delta E^2}{E_0^2}+20\frac{\Delta E^3}{E_0^3}-15\frac{\Delta E^4}{E_0^4}-6\frac{\Delta E^5}{E_0^5}-\frac{\Delta E^6}{E_0^6}\right) 
\end{equation}
where we have used the approximation $(1-e^{-A_\beta \tau_R})\approx A_\beta \tau_R=f_{pp}$ because
in all interesting experimental configurations $f_{pp}\ll0.01$.

Substituting (\ref{eq:F0}), (\ref{eq:Fm}) and (\ref{eq:frac_pp_re}) in (\ref{eq:S2Neq}), keeping only the terms up to 
$(\Delta E/E_0)^3$ and considering that $(1+f_{pp})\approx1$, 
we obtain
\begin{equation}\label{eq:sens_eq_2}
 \frac{m_\nu^2}{E_0^3}\left( \frac{3}{2}\Delta E - \frac{3 m_\nu^2}{8\Delta E} \right) \sqrt{ N_{ev}} =
 1.7 \sqrt{\frac{\Delta E^3}{E_0^3} + f_{pp}  \left(\frac{3\Delta E}{10E_0} + \frac{3 \Delta E^2}{4 E_0^2}\right)+b\Delta \! E/A_\beta}
\end{equation}
which can be solved for $m_\nu$ to give the sensitivity at 90\% confidence level, $\Sigma_{90}(m_\nu)$.
Considering only the leading terms in (\ref{eq:Fm}) and (\ref{eq:frac_pp_re}) then the solution is
\begin{equation}
\label{eq:sensitivity}
\Sigma_{90}(m_\nu) = 1.13  \frac{E_0}{\sqrt[4]{N_{ev}}} \left[{\frac{ \Delta E}{E_0} + \frac{E_0}{\Delta E} \left( \frac{3}{10} \ f_{pp}
+b \frac{E_0}{A_\beta}\right)}\right]^{\frac{1}{4}}
\end{equation}
In order to make meaningful use of  (\ref{eq:sensitivity}) one has to interpret correctly the energy interval $ \Delta E$. The same applies to the solution of  (\ref{eq:sens_eq_2}).

The two terms in the square bracket in (\ref{eq:sensitivity}) represent  the contributions to the {\it noise}
from the statistical fluctuations  of the beta and pile-up spectra respectively (here we neglect the background term for sake
of clarity).
When the pile-up term is negligible (because of the low rate $A_\beta$ or of the short resolving time $\tau_R$) the left term dominates and  it pays out to
keep $\Delta E$ as small as possible: the limit is of course the detector energy resolution.
On the other extreme, when the end-point of the beta spectrum is buried in the pile-up spectrum, the {\it noise} is dominated by the right term.
In this case the signal-to-noise ratio improves by enlarging the energy interval $ \Delta E$.

It is then clear that there is no defined value of  $\Delta E$ to plug in  (\ref{eq:sensitivity}): the solution we have found is to choose $\Delta E$ as the 
value that minimizes $\Sigma_{90}(m_\nu)$ for a given set of experimental parameters, with the boundary condition that $\Delta E$ cannot be smaller than the detector energy resolution \de\fwhm.

In particular, for the simpler case of  (\ref{eq:sensitivity}), by 
searching the positive zero of the derivative with respect to \de, we can obtain the following
\begin{equation}
\label{eq:de_opt}
 \Delta E = max\left(E_0\sqrt{\frac{3}{10}f_{pp}+b \frac{E_0}{A_\beta}},\,\, \Delta E_\mathrm{FWHM}\right) 
\end{equation}

This approach for defining \de\ can be applied to the solution of  (\ref{eq:sens_eq_2}) as well.
In this case \de\ can be evaluated numerically following the above prescriptions. In particular to obtain the results presented in this paper, after making the substitution $\Delta E \rightarrow |\Delta E_{opt}| + \Delta E_\mathrm{FWHM}$,  we have found numerically the $\Delta E_{opt}$ which
minimizes the solution of (\ref{eq:sens_eq_2})
\begin{equation}
\label{eq:de_opt_2}
 \Sigma_{90}(m_\nu) = f(\Delta E_{opt}, \Delta E_\mathrm{FWHM}, \tau_R, A_\beta, N_{det}, t_M,b) 
\end{equation}
It is worth noting that,  in this analysis, equations (\ref{eq:de_opt}) and (\ref{eq:de_opt_2}) are the only places where the
detector energy resolution $\Delta E_\mathrm{FWHM}$ shows up.

\subsection{Montecarlo approach}
In this section we describe  a frequentist Montecarlo code developed to estimate the statistical sensitivity of a neutrino mass
experiment performed with thermal calorimeters. 
The approach is to simulate the beta spectra that would be measured by 
a large number of experiments carried out in a given configuration: the spectra are then fit 
as the real ones \cite{MIBETA} and the statistical sensitivity is deduced from the distribution of the 
obtained $m^2_\nu$ parameters.

The Montecarlo parameters describing the experimental configuration are the total statistics $N_{ev}$,
the FWHM of the Gaussian energy resolution \de \fwhm, the fraction of unresolved pile-up events $f_{pp}$ and
the background $B(E)$.
These input parameters can be derived from the ones actually characterizing a real experiment: 
$N_{ev} = N_{det}  A_{\beta}  t_M$ and, by recalling (\ref{eq:frac_pp}), \mbox{$f_{pp} = F^{pp}_{\Delta E = E_0}\approx A_\beta \tau_R$}, where again
$N_{det}$ is the number of detectors, $A_{\beta}$ is the beta decay activity of a single detector, 
$t_M$ is the measuring time and $\tau_R$ is the pile-up resolving time.   

The procedure to estimate the statistical sensitivity goes through the following steps:
\begin{itemize}
 \item The theoretical spectrum $S(E)$ which is expected to be measured by the virtual experiments is evaluated:
\begin{equation}
\label{eq:modelMC}
S(E) = \left[N_{ev}(N_\beta(E,m_\nu)+f_{pp}N_\beta(E,0)\otimes N_\beta(E,0))+B(E)\right]\otimes R(E)
\end{equation}
where $N_\beta(E,m_\nu)$ is again the  $^{187}$Re beta spectrum normalized to unity (\ref{eq:quad}), $B(E)$ the background
energy spectrum and $R(E)$ is the detector energy response function. The $B(E)$ function is usually taken as a constant $B(E)=bT$.
The response function $R(E)$  is assumed to be a symmetric Gaussian 
\begin{equation}
\label{eq:gauss}
 G(E)=\frac{1}{\sigma \sqrt{2\pi}}e^{-\frac{E^2}{2\sigma^2}}
\end{equation}
with standard deviation $\sigma = \Delta E_{\mathrm {FWHM}}/2.35$.
\item The virtual outcome of a large number (between 100 and 1000) of experiments is numerically generated by 
letting the spectrum $S(E)$ fluctuate according to a Poisson statistics. The simulated experimental spectra are generated on an energy interval which is smaller than the full 0 -- $2 E_0$ interval.
\item Each simulated spectrum is fitted using 
(\ref{eq:modelMC}) and leaving $m^2_\nu$, $E_0$, $N_{ev}$, $f_{pp}$ and $b$ as
free paramaters. The fit is restricted to an energy interval smaller than the one used for the simulated spectrum generation.
\item The 90\% C.L. $m_\nu$ statistical sensitivity $\Sigma_{90}(m_\nu)$ of the simulated experimental 
configuration is given by $\Sigma_{90}(m_\nu) = \sqrt{1.7 \sigma_{m_\nu^2}}$, where $\sigma_{m_\nu^2}$ is the standard deviation of the distribution of the $m^2_\nu$ found by fitting the spectra.
\begin{equation}
\sigma_{m_\nu^2}^2 = \frac{1}{N-1} \sum_i (m_{\nu_i}^2 - \overline{m_\nu^2})^2 = \frac{N}{N-1} (\overline{m^4_\nu} - \overline{m^2_\nu}^2)
\end{equation}
where $N$ is the number of generated spectra and $m_{\nu_i}^2$ are the values found in each fit for $m^2_\nu$ fit parameter.
\item The statistical error on the 90\% C.L. $m_\nu$ statistical sensitivity is estimated as follows.
By defining $y_i=(m_{\nu_i}^2 - \overline{m_\nu^2})^2$, we have  $\overline{y} \approx \sigma_{m_\nu^2}^2$ and we can write
\begin{equation}
\sigma^2_y = \frac{N}{N-1} (\overline{y^2} - \overline{y}^2) \approx \frac{N}{N-1} \left[ \frac{1}{N} \sum_i(m_{\nu_i}^2 - \overline{m^2_\nu})^4 -  \sigma_{m_\nu^2}^4 \right]
\end{equation}
The error on $\sqrt{\overline y} =\sigma_{m_\nu^2}$ is given by 
\begin{equation}
 \epsilon_{\sqrt{\overline y}} = \frac{1}{2}\sqrt{\frac{\sigma^2_y}{N \sigma_{m_\nu^2}} } 
\end{equation}
and therefore the error on $\Sigma_{90}(m_\nu)$ is obtained 
\begin{equation}
\label{eq:errorsigma}
\epsilon_{\Sigma_{90}(m_\nu)} = \frac{1.7}{2} \frac{\epsilon_{\sqrt{\overline y}}}{\Sigma_{90}(m_\nu)}
\end{equation}
Using equation (\ref{eq:errorsigma}) one finds that the statistical error on the Montecarlo results is around 3\% and 1\% for about 100 and 1000 simulated experiments respectively.

\end{itemize}

\subsection{Analytic vs. Montecarlo}
\label{sec:ana-vs-mc}
\begin{figure}[tb!]
\begin{center}
\resizebox{.8\textwidth}{!}{ \includegraphics*[]{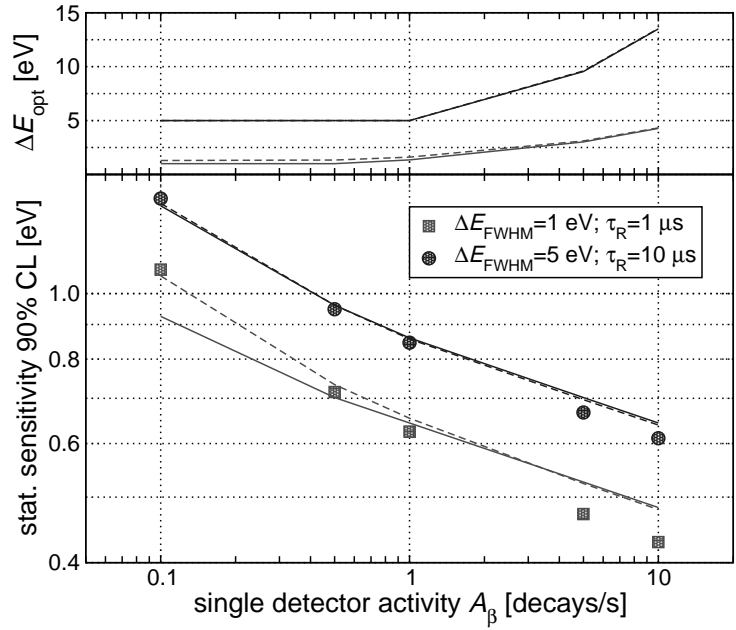}}
\caption{\label{fig:fixT}Comparison between the statistical sensitivity as estimated by a Montecarlo approach (symbols) 
and by the analytic formulation (lines). 
The continuous and dashed lines are obtained using (\ref{eq:sensitivity}) with (\ref{eq:de_opt}) and (\ref{eq:de_opt_2}) respectively. 
The statistical sensitivity is evaluated 
for an exposure $T=10000$\,detector$\times$year.
The rightmost point corresponds to a total statistics $N_{ev}=3\times10^{12}$.
The upper panel shows how $\Delta E_{opt}$ gets larger as the pile-up spectrum increase its weight. 
}
\end{center}\end{figure}

\begin{figure}[tb!]
\begin{center}
\resizebox{.8\textwidth}{!}{ \includegraphics*[]{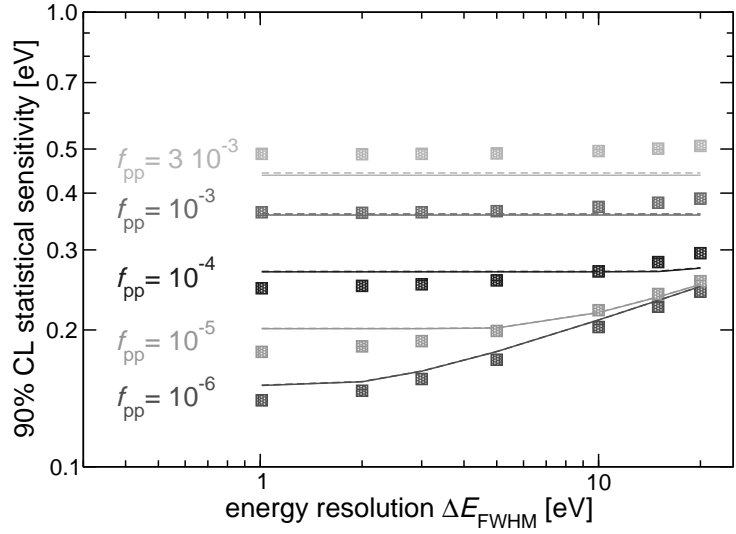}}
  \caption{\label{fig:cfr}  Comparison between the statistical sensitivity as estimated by a Montecarlo approach (symbols) and by the analytic formulation (lines) for a total statistics $N_{ev}$ of $10^{14}$ events. Continuous and dashed lines are as in Figure\,\ref{fig:fixT}.}
\end{center}\end{figure}

\begin{figure}[tb!]
\begin{center}
\resizebox{.8\textwidth}{!}{ \includegraphics*[]{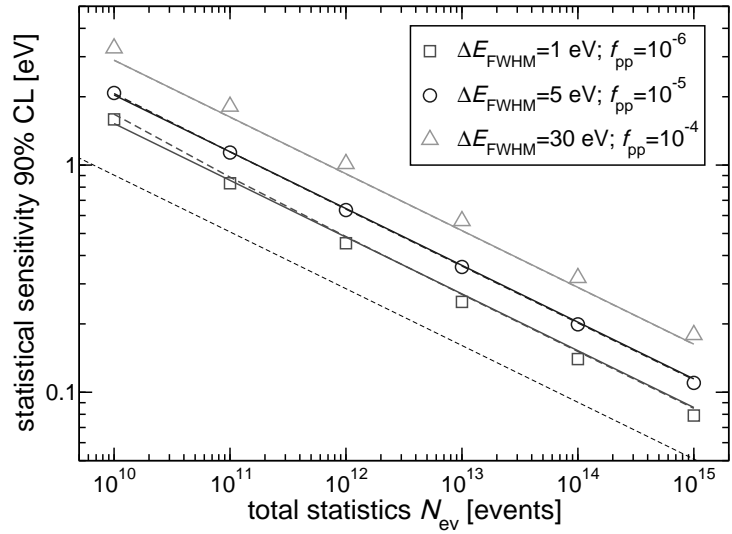}  }
  \caption{\label{fig:scan-Nev} Comparison between the statistical sensitivity as estimated by a Montecarlo approach (symbols) and by the analytic formulation (lines). Continuous and dashed lines are as in Figure\,\ref{fig:fixT}. The fine dashed line shows a $N_{eV}^{1/4}$ functional dependence for sake of comparison.}
\end{center}\end{figure}

\begin{figure}[tb!]
\begin{center}
\resizebox{.8\textwidth}{!}{ \includegraphics*[]{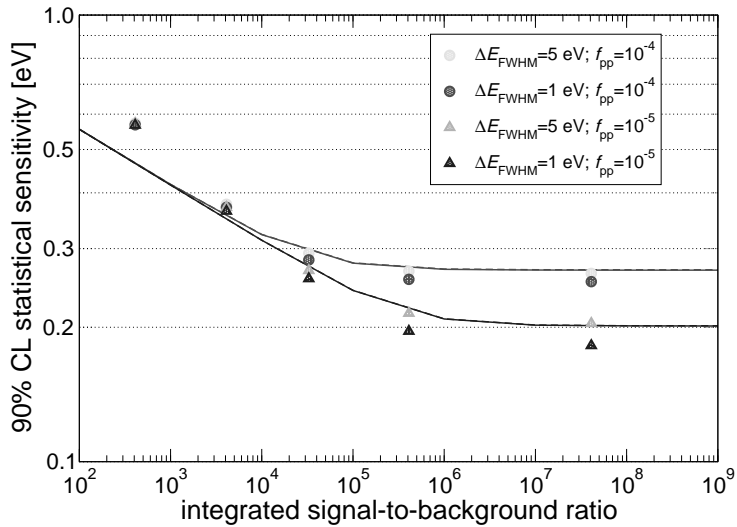}}
  \caption{\label{fig:bkg} Comparison between the statistical sensitivity as estimated by a Montecarlo approach (symbols) and by the analytic formulation (lines). The integrated signal-to-background ratio is given by $N_{ev}/N_{bkg}$ where $N_{bkg}=bE_0T$ (see text) and $N_{ev}=10^{14}$. An integrated ratio of about $3\times10^4$ corresponds to the background level measured in the Milano experiment.}
\end{center}\end{figure}

We have compared the predictions of the two approaches described in the previous sections for a wide range of experimental
configurations suitable for obtaining a sub-eV neutrino mass sensitivity (Figure\,\ref{fig:fixT}, \ref{fig:cfr}, \ref{fig:scan-Nev} and \ref{fig:bkg}). In all plots the continuous lines are obtained through equations (\ref{eq:sensitivity}) and (\ref{eq:de_opt}), while
the dashed lines are obtained from (\ref{eq:sens_eq_2}) as described at the end of the section on the statistical sensitivity.
The symbols are the results of the Montecarlos, which have negligible errors on these scales (see equation (\ref{eq:errorsigma})).

Figure\,\ref{fig:fixT} shows how the sensitivity improves for increasing single detectors activity $A_\beta$, with
the other experimental parameters fixed and for an exposure $T$ of $10000$\,det$\times$year. From this plot one can deduce
that it pays out to increase the single detector activity at the expense of an increased fraction of pile-up events, $f_{pp}$.
The upper panel of Figure\,\ref{fig:fixT} shows the value of the optimal energy interval $\Delta E_{opt}$ defined as discussed above. 

Figure\,\ref{fig:cfr} demonstrates how the impact of the detector FWHM energy resolution \de\ on the sensitivity is reduced 
by a high pile-up fraction $f_{pp}$. This Figure shows also one of the limits of the analytic approach described above, i.e.
the poor consideration of the detector energy resolution which translates in a too weak dependence of the sensitivity on this parameter.

From Figure\,\ref{fig:scan-Nev} it is possible to appreciate the importance of the total statistics $N_{ev}$ to reach a neutrino mass sensitivity of the order of 0.1\,eV. In particular energy resolutions \de\ of about 1\,eV and pile-up fractions of the order
of $10^{-6}$ are required. The fine dashed line on the plot corresponds to a $N_{ev}^{-1/4}$ functional dependence of the sensitivity (recall equation (\ref{eq:sensitivity})): this dependence may be exploited to scale the Montecarlo results.

In Figure\,\ref{fig:bkg} the impact of the continuous background below the beta spectrum is shown. The abscissa is the ratio
between $N_{ev}$ and the total number of background counts $N_{bkg}$ between 0 and $E_0$, i.e. $N_{bkg}=b E_0 T$.
Clearly the impact is lower for higher pile-up fractions $f_{pp}$.

For more details on each plots the reader can refer to their captions. In the final section of this paper these results 
will be used to assess the potential of a calorimetric neutrino mass experiment using \Re.

Although the agreement is only partial, the comparison confirms that the analytic formulation goes in the right direction
to be used to make useful predictions. Nevertheless we believe that the most accurate estimate of the sensitivity is 
the one obtained through the Montecarlo frequentist approach.
From the Figures it is apparent that the formula tends to overestimate the sensitivity for increasing $\Delta E_{opt}$, i.e. when the sensitivity is limited by the pile-up or the continuous background.
To improve the predictive power of the analytical approach it is possible to introduce free parameters in the formulas and to adjust them to better reproduce the Montecarlo predictions, but this is out of the scope of this paper.

As a general check, both approaches can be applied taking as input the experimental parameters of the Milano experiment with an array of \agre\ crystals whose results are presented in \cite{MIBETA}. 
For $N_{ev}=1.7\times10^7$, \de\fwhm\,=\,28.5\,eV, $f_{pp}=2.3\times10^{-4}$ and $N_{ev}/N_{bkg}=3.28\times10^4$ 
one obtains a sensitivity at 90\%\,C.L. of about 15 and 17\,eV, using equations (\ref{eq:sensitivity}) and the Montecarlo respectively, while the limit on the neutrino mass reported in \cite{MIBETA} is about 16\,eV at  90\%\,C.L.

\section{Systematic uncertainties}
\label{sec:sys}
Although, as mentioned in the introduction, a calorimetric neutrino mass experiment is considered free from  systematics
related to the external source effects, still it may be affected by other uncertainties. 

As it will be discussed in the 
following, the origin of some of these effects (electron escape, beta decay spectral shape and beta environmental fine structure) is indeed related to the beta source and may be unavoidable in spite of the calorimetric configuration. In order to minimize the related uncertainties, this kind of effects
must be precisely modelled with the help of
theoretical investigations, independent experiments and Montecarlo simulations.

Other systematic uncertainties arise from instrumental effects and can be mitigated through an improved detector design and characterization as well as a careful off-line data analysis.

The frequentist Montecarlo code described above can be readily adapted to  estimate  the many systematic effects which in general fall under two categories. To the first belong the uncertainties due
to  lack of accuracy with which the experimental parameters are determined. To assess the corresponding systematic uncertainties in the generated spectra,
the parameters are randomly fluctuated -- according to the given accuracy -- while they are kept fixed to their average value
in the fitting function $S(E)$ (see (\ref{eq:modelMC})).
The second category consists in the effects caused by an incomplete or incorrect modelling of the data. In this case the adopted 
approach is to include the effects in the generated spectra, but not in the fitting function $S(E)$. 

In general these procedures result in a shift of $m_\nu^2$ away from zero  
and, in some cases, in a sensible deterioration of the sensitivity as shown by the wider error bars in the plots.
The purpose of this analysis
is to identify the size of the inaccuracy or of the neglected effect for which the shift of $m_\nu^2$ remains within a given limit.
Table\,\ref{tab:sys} reports the results in terms of the systematic uncertainty giving a shift of $|m_\nu^2|$ less than 0.01\,eV$^2$ - value for which the systematic uncertainties remain smaller than the statistical
error of an experiment aiming at a sub-eV sensitivity.

In the following we analyze in more details the effects that are more likely to introduce systematic uncertainties in calorimetric neutrino mass experiments. Except where differently stated, the plots in this section are obtained for $N_{ev} = 10^{14}$, $\Delta E_\mathrm{FWHM}=1.5$\,eV and $f_{pp}=10^{-6}$.

\subsection{Source related uncertainties}
\label{sec:source-sys}
\paragraph{Excited final states}
Beta decays to any kind of excited final state are not going to perturb the beta spectrum end-point as long as the  
state lifetimes are shorter than the detector integration time -- which is always more than about 1\,\mus.
In fact, while an excitation energy $E_{exc}$ lost in the beta decay shifts the beta spectrum end-point to $E_0'=E_0 - E_{exc}<E_0$, at
the same time the coincident detection of the energy  $E_{exc}$ released in the state de-excitation adds an energy offset to
the beta spectrum. The final outcome is that to each excited state corresponds a beta spectrum starting at $E_{exc}$ and with 
end-point $E_0=E_0'+E_{exc}$.
Thanks to its simple functional energy dependency  (\ref{eq:quad}) the \Re\ beta decay spectral shape above $E_{exc}$ remains unperturbed.\footnote{This is not true for a more general beta spectrum shape as for example in the case of allowed transitions with $N(E)\propto p E (E_0-E)^2 F(Z,E)$, where $p$ is the electron momentum and $F(Z,E)$ is the Fermi factor. For calorimeters, it is more generally true that at the end-point
the correction for the presence of excited final states vanishes approximately as $\sum_i (1+V_i/E)$, where $V_i$ are the final state energies.}

\begin{figure}[bth!]
\begin{center}
\resizebox{.8\textwidth}{!}{ \includegraphics*[]{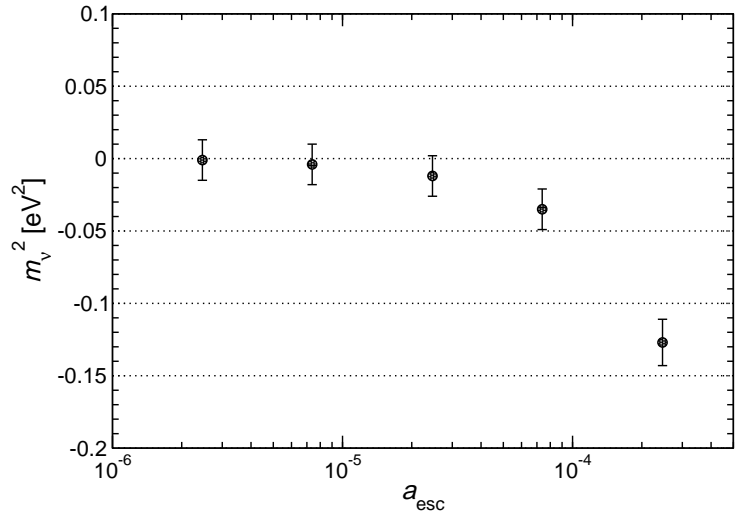}}
  \caption{\label{fig:esc} Systematic \mnsq\ shift caused by ignoring the correction for the surface electron
escape. Points obtained for $N_{ev} = 10^{14}$, $\Delta E_\mathrm{FWHM}=1.5$\,eV, $f_{pp}=10^{-6}$ and $bT=0$\,c/eV.
}
\end{center}\end{figure}

\paragraph{Electron escape}
A  fraction of electrons emitted in the decays of \Re\ nuclei close to the detector surface will not be contained
in the calorimeter. We have used the Geant4 toolkit \cite{Geant4} to estimate the type and magnitude of this effect on the measured spectrum.
\begin{figure}[tbh!]
\begin{center}
\resizebox{.8\textwidth}{!}{ \includegraphics*[]{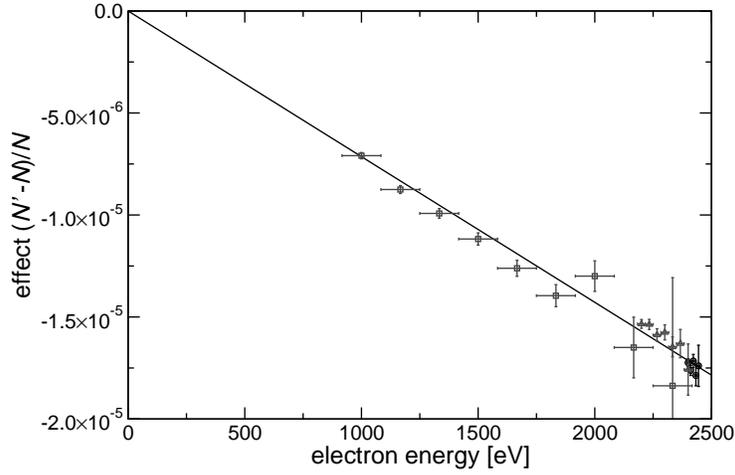}}
  \caption{\label{fig:escape} Spectral distortion caused by surface electron escape as calculated by
a Montecarlo simulation for a 1\,mg cubic Rhenium detector. The effect plotted in the graph is  $(N'(E) - N(E))/N(E)$, where $N(E)$ is given by equation (\ref{eq:quad}) and $N'(E)=N(E)f_{esc}(E)$}
\end{center}\end{figure}
Figure\,\ref{fig:escape} shows the results for a 1\,mg cubic Rhenium detector - i.e. with linear dimensions of about
0.362\,mm -  in terms of relative deviation with respect to the spectrum given by (\ref{eq:quad}).
The simulation has been repeated for the two available Geant4 low energy extensions (the results in Figure\,\ref{fig:escape} are the ones obtained using the Penelope extension) and for different low energy cuts applied in the electron transport. These tests has confirmed the 
shape of the effect and its magnitude, while giving slightly different results. In conclusion the Montecarlo simulation cannot
be considered reliable to precisely calculate the effect, also considering the uncertainties in the shape and size of the 
detector rhenium absorber. 
The effect on the measured spectrum can be parametrized as a multiplicative factor to include in (\ref{eq:quad}) given by
\begin{equation}
\label{eq:esc}
f_{esc}(E) = 1 - a_{esc} \frac{E}{E_0}
\end{equation}
where the dimensionless $a_{esc}$ parameter will have to be left free in the data analysis. 
The solid line in Figure\,\ref{fig:escape} corresponds to $a_{esc} = 1.9\times 10^{-5}$.
We estimated the systematic error arising when this effect is not included in the data analysis for various values
of $a_{esc}$. The results are plotted in Figure\,\ref{fig:esc}.

\subsection{Beta spectrum uncertainties}
\begin{figure}[htb!]
\begin{center}
\resizebox{.95\textwidth}{!}{ \includegraphics[bb=50 68 410 280, clip=true]{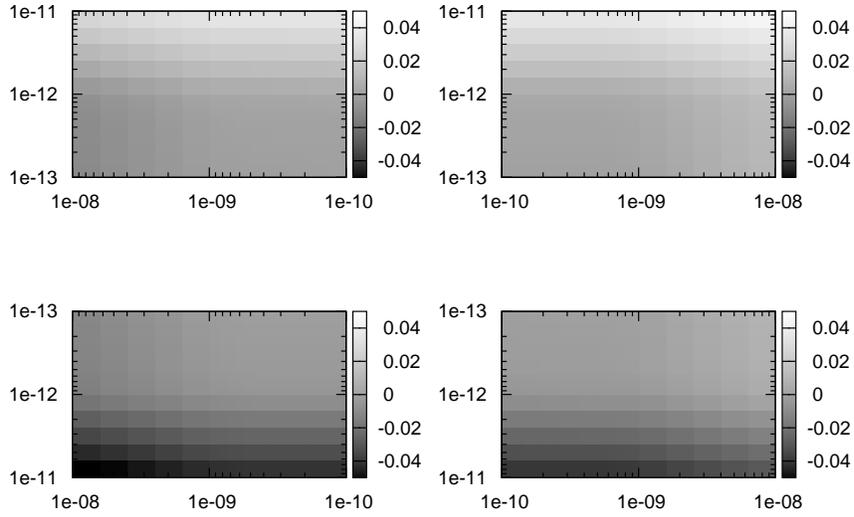}}
  \caption{\label{fig:spemulti} Systematic \mnsq\ shift due to a deviation of the beta spectrum shape from
the simple quadratic form in (\ref{eq:quad}).
The four graphs represent the four quadrants of the $(a_1,a_2)$ plane, where $a_1$ and $a_2$
are the correction coefficients introduced in (\ref{eq:corr_spe}). Starting from upper right graph
and going clockwise, they are the $a_1$\,vs.\,$a_2$, $a_1$\,vs.\,$-a_2$, $-a_1$\,vs.\,$-a_2$, and  $-a_1$\,vs.\,$a_2$ plots. The colour coded z-axis is the $m^2_\nu$ value. The Montecarlo parameters are $N_{ev} = 10^{14}$, $\Delta E_\mathrm{FWHM}=1.5$\,eV $f_{pp}=10^{-6}$ and $bT=0$\,c/eV.}
\end{center}\end{figure}

\begin{figure}[htb!]
\begin{center}
\resizebox{.95\textwidth}{!}{ \includegraphics[bb=70 68 400 260, clip=true]{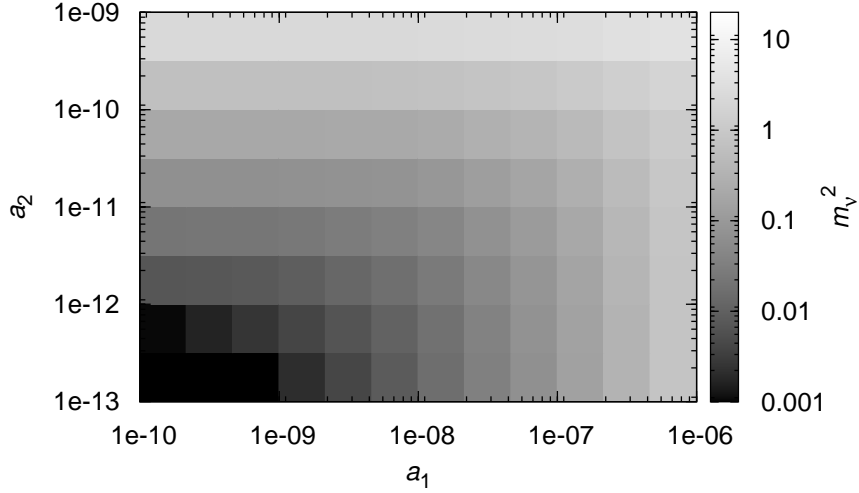}}
  \caption{\label{fig:spe} Detail of the $a_1>0$\,vs.\,$a_2>0$ quadrant for a wider parameter range and with
logarithmic z-axis.}
\end{center}\end{figure}

\paragraph{Spectral shape}
Although the use of equation (\ref{eq:quad}) for the \Re\ beta decay spectrum is up to the purpose of the present
work, future high statistic experiments will need a more precise description of the spectrum. In order to estimate
the sensitivity to deviations from the simple equation  (\ref{eq:quad}) we have considered the corrective factor 
\begin{equation}
\label{eq:corr_spe}
f_{corr}(E) = 1+a_1 E+a_2 E^2
\end{equation}
which is actually an extension of the correction for the escape of beta electrons (\ref{eq:esc}).
The effect of deviations of this kind has been investigated by generating the experimental spectra according to 
a modified beta spectrum $N'(E)=N(E)f_{corr}(E)$ and fitting them using the regular beta spectrum $N(E)$ in $S(E)$ (\ref{eq:modelMC}).
Figure\,\ref{fig:spemulti} shows the effect on $m_\nu^2$ varying the $a_i$ coefficients (see the caption for more details). 
Figure\,\ref{fig:spe} displays the results for positive values of the  $a_i$ coefficients.

\begin{figure}[htb!]
\begin{center}
\resizebox{.8\textwidth}{!}{ \includegraphics*[]{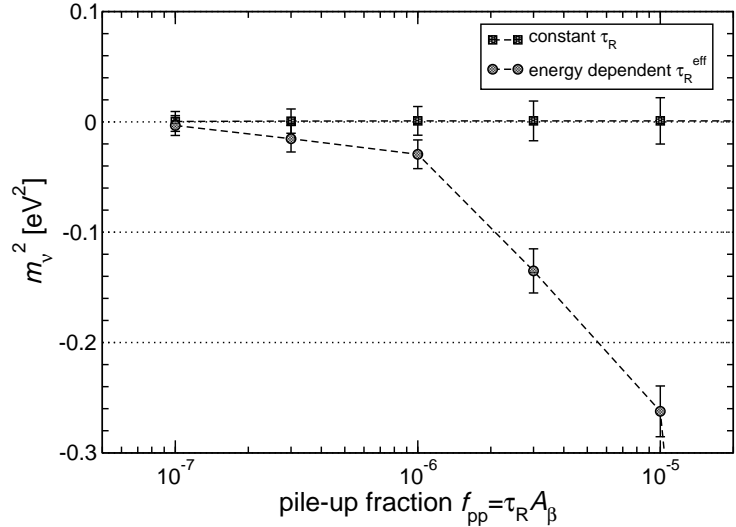}}
  \caption{\label{fig:syspp} Systematic \mnsq\ shift caused by ignoring the pile-up spectrum correction for an energy dependent pile-up rejection efficiency. Points obtained for $N_{ev} = 10^{14}$, $\Delta E_\mathrm{FWHM}=1.5$\,eV, and $bT=0$\,c/eV.}
\end{center}\end{figure}

For sake of comparison, one can consider that the deviation of the approximate beta spectrum shape described by equation (\ref{eq:quad}) 
from the theoretical one given in \cite{simkovic} can be parametrized as
\begin{equation}
f_{corr}(E) \approx 1.0 - 1.8\times10^{-5} E + 2.8\times10^{-10} E^2 - 3.5\times10^{-15} E^3 + ...
\end{equation}
Neglecting such a correction in the data analysis would systematically shift   $m^2_\nu$ by about -280\,eV$^2$. 
\paragraph{Beta Environmental Fine Structure}
\begin{figure}[htb!]
\begin{center}
\resizebox{.8\textwidth}{!}{ \includegraphics*[]{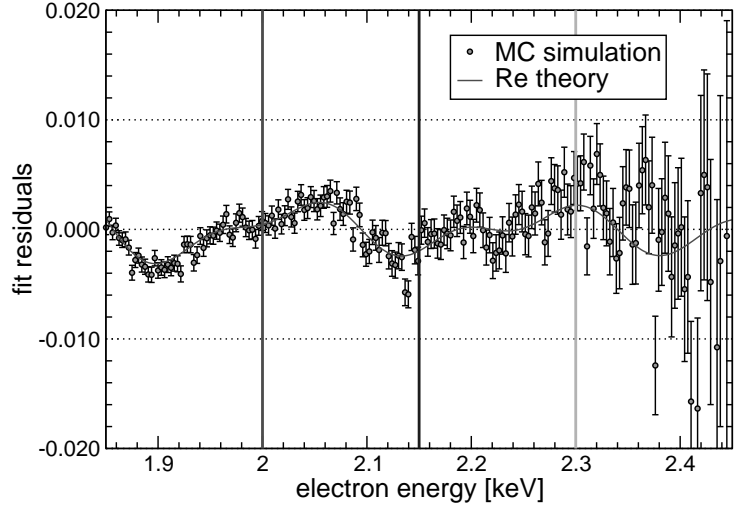}}
  \caption{\label{fig:befs} Residuals from the fit of a Montecarlo generated beta spectrum with BEFS using a fit function without BEFS.
The Montecarlo is for $N_{ev}=10^{10}$, $\Delta E_\mathrm{FWHM} = 5$\,eV, $f_{pp}=10^{-5}$  and $bT=0$\,c/eV.}
\end{center}\end{figure}
The Beta Environmental Fine Structure (BEFS) is a modulation of the beta emission probability due to the
atomic and molecular surrounding of decaying nuclei: it is the analogous of the oscillation observed in the Extended X-ray Absorption
Analysis (EXAFS) and it is explained by the electron wave structure in terms of reflection and interference.
Although the phenomenon is completely understood, its description is quite complex and the parameters involved are not known a-priori.
Because of its faintness, so far the BEFS has been observed in metallic Rhenium \cite{Ge-Nature} and in \agre\ \cite{BEFS-PRL} only below 1.5\,keV where it
is larger. 
It is clear that future neutrino mass experiments will cope with the need of a very accurate description of the BEFS modulation up to the
beta spectrum end-point.
The parameters presently available are still affected by large statistical errors: for a safe extrapolation up to the end-point the
BEFS must be characterized using much higher statistics spectra. 
Meanwhile the Montecarlo approach can be used to show the  shift on $m_\nu^2$ when data with BEFS included are fitted to a model without  BEFS. For the BEFS function it is assumed that the one used to interpolate the data up to 1.5\,keV can be used up to the end-point without modifications. 
A Montecarlo simulation of the Rhenium BEFS in a measurement with a statistics of about $10^{10}$ events is shown in Figure\,\ref{fig:befs} in terms
of residuals of the fit.
Figure\,\ref{fig:befs_sys} shows what happens to $m_\nu^2$ when fitting spectra like the one in Figure\,\ref{fig:befs} with different left boundaries of the fitting energy interval. 
The effect worsens when the left boundary is moved to lower energies where the BEFS gets larger. The plot confirms that the inclusion of the BEFS in the end-point analysis is mandatory.
\begin{figure}[htb!]
\begin{center}
\resizebox{.8\textwidth}{!}{ \includegraphics*[]{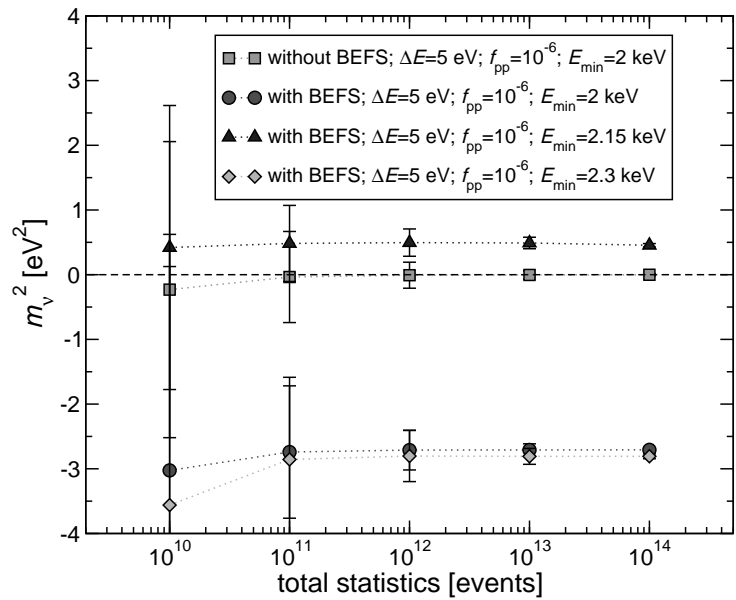}}
  \caption{\label{fig:befs_sys} Systematic effect caused on \mnsq\ by neglecting the BEFS when fitting the
Montecarlo generated spectra with different left energy boundaries (see Figure\,\ref{fig:befs}).}
\end{center}\end{figure}
\paragraph{Pile-up spectrum}
As discussed in \S\,\ref{sec:analytic}, formula (\ref{eq:npp}) holds only under the assumption 
of a constant resolving time $\tau_R$. For real detectors  $\tau_R$ depends on the
pulse shape and on the noise level: in practice $\tau_R$ tends to increase  for smaller pulses.
The detailed behavior of the resolving time is difficult to predict and must be modelled
by Montecarlo methods taking in account both the actual pulse and noise frequency 
spectra and the algorithm used to identify the double pulses. Examples of such analysis can 
be found in \cite{MANU-pup, MARE-pup}.
In order to get a sensible idea of the systematics related to this effect, we have used the results presented in \cite{MANU-pup}. 
We have numerically evaluated the pile-up spectrum introducing a variable pile-up 
rejection efficiency described by an effective resolving time $\tau_R^{\mathit{eff}}$ \begin{equation}
\label{eq:taureff}
\tau_R^{\mathit{eff}} = \tau_R \left[1 + (n_{\tau_R} - 1) e^{-r_A\lambda_r}\right]
 \end{equation}
where $r_A<1$ is the ratio between the amplitudes of the two events to discriminate, $n_{\tau_R}$ is 2 and $\lambda_r$ is 4.0. The function (\ref{eq:taureff}) roughly approximates the one described in  \cite{MANU-pup}, even though it neglects the time ordering of the two events. 
With this approximation we find about 30\% more unresolved pile-up events and a good description
of the resulting pile-up spectrum for $A_\beta \tau_R\ll 0.1$ is given by
\begin{equation}
N^{\prime}_{pp}(E)=(1-e^{-A_\beta \tau_R})N_\beta(E,0) \otimes N_\beta(E,0) \left(1 + \frac{0.35}{e^{(E-E_0)/(480.0\,\mathrm{eV})} + 1}\right)
\end{equation}
Figure\,\ref{fig:syspp} shows that the systematic shift caused by neglecting this deviation
increases with the pile-up probability $\tau_R A_\beta$. In particular it is apparent as 
a proper modelling of the pile-up spectrum is crucial for a pile-up rate as low as $10^{-6}$.

\subsection{Instrumental uncertainties}
\label{sec:instr-sys}
\begin{figure}[htb!]
\begin{center}
\resizebox{.8\textwidth}{!}{ \includegraphics*[]{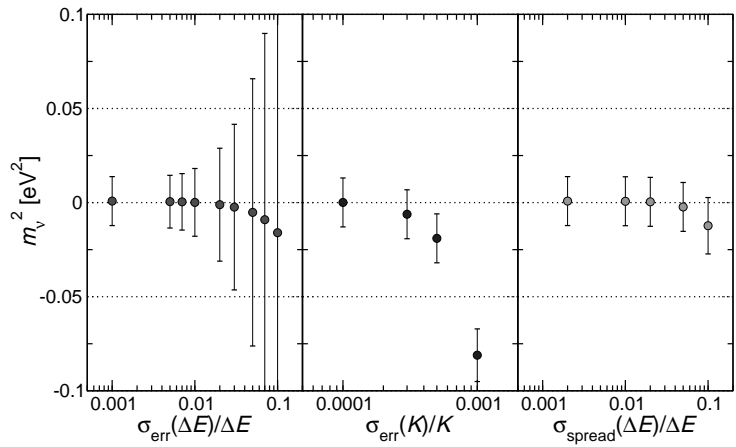}}
  \caption{\label{fig:vari} Instrumental systematic uncertainties for $N_{ev}=10^{14}$, $\Delta E_\mathrm{FWHM} = 1.5$\,eV, $f_{pp}=10^{-6}$  and $bT=0$\,c/eV: response function uncertainty (left), energy calibration errors in an array (center) and detector energy resolution spread in an array (right). }
\end{center}\end{figure}
\paragraph{Response function uncertainty}
In (\ref{eq:modelMC}) the simplest response function $R(E)$ used to model the data is a Gaussian $G(E)$ (\ref{eq:gauss}) which
is completely determined by its standard deviation $\sigma=\Delta E/2.35$.
The detector FWHM energy resolution \de\ is usually determined by means of a calibration procedure using radioactive sources. The accuracy with which \de, and therefore $R(E)$, is known is mainly limited by statistics.

Assuming a purely Gaussian $R(E)$,
the systematics due to the finite accuracy with which the detector FWHM energy resolution \de\ is known have been evaluated by letting fluctuate 
the detector energy resolution in the simulated spectra around a central value  \de\ which is the fixed resolution used to fit the spectra.
The detector energy resolution fluctuates according to a Gaussian distribution  centered in \de\ with standard deviation $\sigma_{err}(\Delta E) $. The resulting shift of $m_\nu^2$ is shown in the left panel of Figure\,\ref{fig:vari}.

\begin{figure}[htb!]
\begin{center}
\resizebox{.8\textwidth}{!}{ \includegraphics*[]{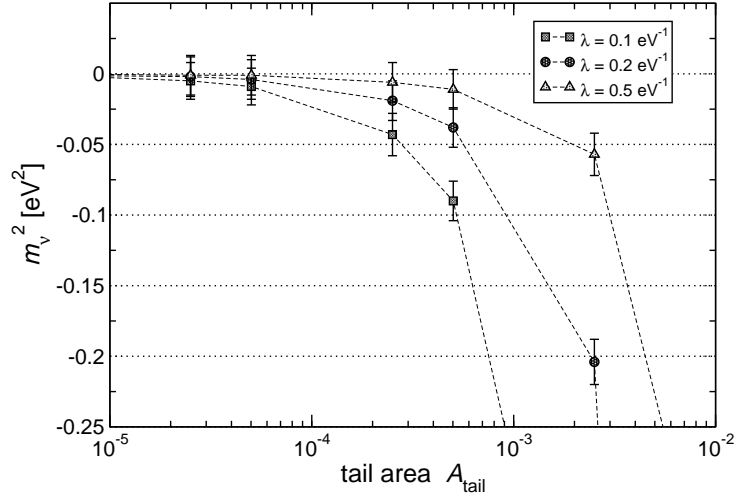}}
  \caption{\label{fig:tail} Systematic effect caused on \mnsq\ by an undetected exponential tail in a Gaussian 
response function $R(E)$. The Montecarlo parameters are $N_{ev}=10^{14}$, $\Delta E_\mathrm{FWHM} = 1.5$\,eV, $f_{pp}=10^{-6}$ and $bT=0$\,c/eV.}
\end{center}\end{figure}

With the number of events $N_{ev}$ and the pile-up fraction $f_{pp}$ considered in the present analysis, a calibration peak at an energy
just above the beta decay end-point $E_0$ would have only the pile-up spectrum as background. In these conditions, a perfectly Gaussian peak with 10$^4$ counts would allow an estimation of the FWHM energy resolution \de\ with an accuracy of about 1\%.
\paragraph{Response function tails}
The actual response function $R(E)$ may be as simple as a Gaussian, though presenting additional extra features which are difficult to identify in the calibration peaks. One example are small tails on the left side of the main Gaussian peak. 
For a Gaussian with variance $\sigma$, the function
\begin{equation}
T(E)  =  A_{tail} \frac{\lambda}{2} \exp \left[ (E - E_{0}) \lambda + \left( \frac{\sigma \lambda}{\sqrt{2}} \right)^2 \right]\left[ 1 - \mathrm{erf} \left( \frac{E - E_{0}}{\sigma \sqrt{2}} + \frac{\sigma \lambda}{\sqrt{2}}\right) \right] 
\end{equation}
represents an exponential tail with area $A_{tail}$ and decay constant $\lambda$.
The effect of such an exponential tail has been studied using a response function $R(E)=G(E)+T(E)$ in $S(E)$ when generating the experimental
spectra, while keeping the standard Gaussian response function in the fit: 
Figure\,\ref{fig:tail} shows the effect for various values of $\lambda$ and $A_{tail}<1$. 

Of course, identifying an exponential tail with a relative area $A_{tail}$ as small as $10^{-4}$ requires a main Gaussian peak with statistics much larger than $10^4$ counts.

\paragraph{Calibration error in array}
Future calorimetric experiments will be carried out with large arrays of thermal detectors ($>10^4$ channels). 
Since each individual channel has to be energy calibrated prior to be summed up, a systematic effect may arise because of the finite calibration accuracy. 
For this analysis each simulated experimental spectrum is the sum of 10000 slightly mis-calibrated spectra. The mis-calibration is simulated shifting the energy calibration according to a Gaussian distribution  centered in $K$ with standard deviation $\sigma(K)$, where $K$ is the correct calibration factor. The center panel of Figure\,\ref{fig:vari} displays the effect on $m_\nu^2$.

In the simplest hypothesis of a linear energy calibration without pedestal, a couple of calibration peaks close to beta end-point
with about $10^4$ counts each and with a background arising solely from the pile-up spectrum would in principle allow to 
determine the calibration factor with an accuracy better than 0.1\%.
\paragraph{Response function dispersion in array}
A second effect that can be observed when summing up many channels is due to the spread in the Gaussian response functions of the single
detectors: the response function of the sum spectrum will not be Gaussian. 
In this case the simulated spectrum is the sum of 10000 ones
whose Gaussian response functions have FWHMs varying according to a Gaussian distribution centered in $\Delta E$ with standard deviation $\sigma_{spread}(\Delta E)$. 
The sum spectrum is analyzed assuming a response function with FWHM equal to $\Delta E$. The results is shown in the left panel of Figure\,\ref{fig:vari}.\footnote{A similar effect due the spread in the resolving time $\tau_R$ is negligible since it affects only the total pile-up spectrum normalization.}

\begin{figure}[tb!]
\begin{center}
\resizebox{.8\textwidth}{!}{ \includegraphics*[]{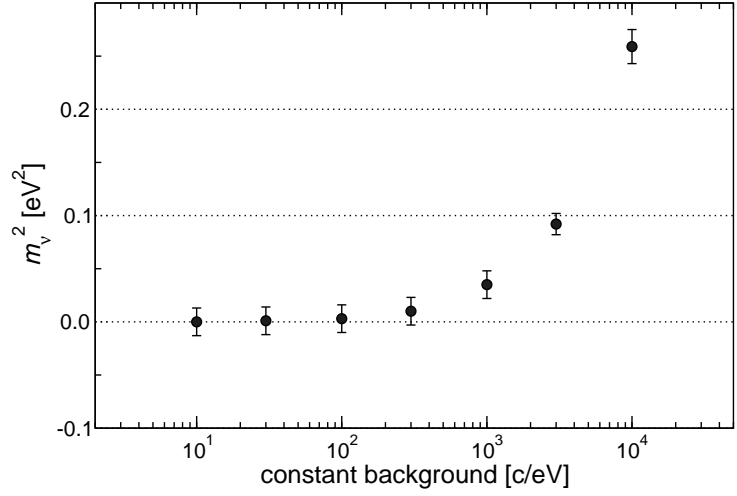}}
  \caption{\label{fig:hidbkg}Systematic shift caused on \mnsq\ by an undetected constant background $bT$ under the beta spectrum. The Montecarlo parameters are $N_{ev}=10^{14}$, $\Delta E_\mathrm{FWHM} = 1.5$\,eV and $f_{pp}=10^{-6}$.
The Milano experiment signal-to-background ratio of about $3\times10^4$ would translates in to 
a constant background $bT$ of about $10^6$\,c/eV on this plot.} 
\end{center}\end{figure}

The dispersion of the energy resolution in an array strictly depends on the detector technology and it is therefore difficult to predict. Nevertheless, based on the experience with running arrays of thermal detectors,
a $\sigma_{spread}$ better than 10\% should be realistic. 
\paragraph{Hidden background} In calorimetric experiments, since the beta source cannot be switched off, the background in the energy
range of the beta spectrum cannot directly assessed. Therefore a costant background is usually included in the fit model $S(E)$ as the safest hypothesis. Nevertheless we have analyzed the effect of neglecting this term. Figure\,\ref{fig:hidbkg} shows
the effect for various levels of constant background $bT$ and confirms the importance of including the background term in $S(E)$.

If no specific measure will be taken to reduce the background, future experiments are expected to have a signal-to-noise
ratio similar to the one of the Milano experiment, i.e. about 10$^4$--10$^5$. In fact increasing the size of the single
detector is unlikely to improve much the ratio since the background is expected to scale approximately as the detector mass.

\begin{figure}[h!]
\begin{center}
\resizebox{.8\textwidth}{!}{ \includegraphics*{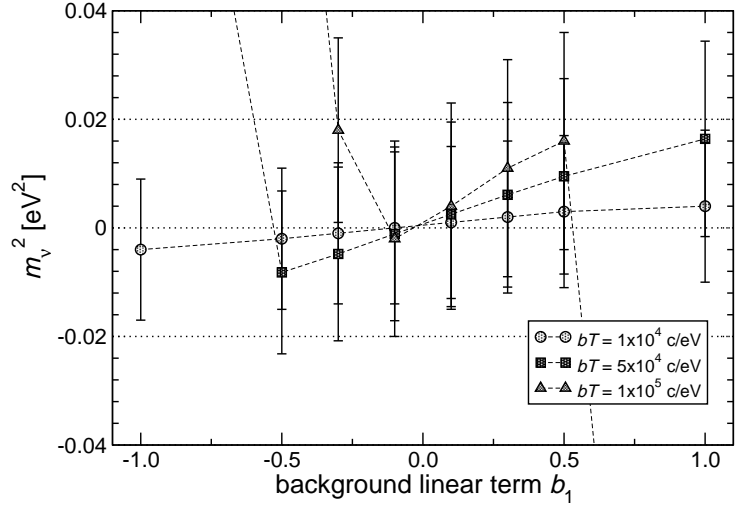}}
  \caption{\label{fig:hidbkg_lin} Systematic shift caused on \mnsq\ by an undetected deviation from flatness of the background under the beta spectrum. The Montecarlo parameters are $N_{ev}=10^{14}$, $\Delta E_\mathrm{FWHM} = 1.5$\,eV and $f_{pp}=10^{-6}$. The Milano experiment signal-to-background ratio corresponds to 
a constant background $bT$ of about $10^6$\,c/eV, therefore one order of magnitude higher than the worst constant background considered in this analysis.}
\end{center}\end{figure}

Eventually the background hidden below the beta spectrum could be not flat.
We have explored this more critical situation making the simple hypothesis of a linear deviation from flatness starting just on the left
of the pile-up spectrum end-point at $2E_0$, expressed as
\begin{equation}
B(E) = bT\left( 1 + \frac{b_1}{E_0}(2E_0 - E) \right)
\end{equation}
The experimental spectra generated with the above linear background were fit with only the constant term in $S(E)$: the results are
shown in Figure\,\ref{fig:hidbkg_lin} for various values of constant background $bT$ and various values for the deviation from flatness $b_1$.

\begin{table}[htb!]
\begin{center}
\renewcommand{\arraystretch}{1.1}
\caption{\label{tab:sys}Analysis of systematic uncertainties for an experiment with $N_{ev}=10^{14}$\,events, $\Delta E=1.5$\,eV and $f_{pp}=10^{-6}$: for each uncertainty, the upper value 
which keeps the systematic shift of $m_\nu^2$ smaller than 0.01\,eV$^2$ is given.}
\begin{tabular}[c]{ccc}
\hline 
\bf source of   & \bf quantity   &\bf maximum effect  \\ 
\bf the uncertainty  &\bf  describing & \bf for  \\ 
\bf   &\bf  the effect & \bf  $\Delta m_\nu^2<0.01$\,eV$^2$  \\ 
\hline 
electron escape & $a_{esc}$ & $1\times10^{-5}$ \\
\hline 

correction to & $|a_1|$ ($a_2=0$) & $\approx10^{-9}$\,eV$^{-1}$\\

theoretical spectral shape & $|a_2|$ ($a_1=0$) & $\approx10^{-12}$\,eV$^{-2}$\\
\hline 

\parbox{0.4\textwidth}{\centering error on energy resolution $\Delta E$}  & ${\sigma_{err}(\Delta E)}/{\Delta E}$  &  0.02 \\ 
\hline 

\parbox{0.4\textwidth}{\centering tail in response function ($\lambda=0.2$\,eV$^{-1}$) } & $A_{tail}$ & $1\times10^{-4}$\\
\hline 

\parbox{0.4\textwidth}{\centering error on single channel energy calibration $K$  } &  ${\sigma(K)}/{K}$   &  $4\times10^{-4}$\\ 
\hline 

\parbox{0.4\textwidth}{\centering spread on energy resolution $\Delta E$ in the array  }  & ${\sigma_{spread}(\Delta E)}/{\Delta E}$  &  0.1\\ 
\hline 

\parbox{0.4\textwidth}{\centering hidden costant background  } & $N_{ev}/N_{bkg}$ & $1\times10^8$\\
\hline 

\parbox{0.4\textwidth}{\centering hidden background linear deviation  ($bT=10^5$\,c/eV)  } & $b_1$ & $\approx 0.1$\\
\hline 
\end{tabular} 
\end{center}
\end{table}

\section{Future calorimetric experiments}
\label{sec:future}
Given that the single channel activity $A_\beta$ is limited by technical considerations concerning the performance of the thermal detector (heat capacity, quasi-particle diffusion length, ...) the question is whether it is desirable to keep the pile-up negligible or not.
There is no unique answer, although increasing the pile-up by increasing $A_\beta$ allows to accumulate more quickly large statistics, and, when pile-up dominates, the dependence on the energy resolution - which tends to degrade when  $A_\beta$ increases -  is attenuated (see Figure\,\ref{fig:fixT} and \ref{fig:cfr}).
On the other hand the  background caused at the end-point by the pile-up, together with a degraded energy resolution, may impair the ability to  recognize and understand systematic effects.
As a conclusion the optimal design of a neutrino mass experiment depends on the detection technique and, in particular, it depends on the
effect of large absorbers on the detector performance. Nevertheless it may pay out to increase the single channel activity as much as possible, therefore relaxing the need of a high energy resolution. 

As an example, Table\,\ref{tab:exp1} and \ref{tab:exp2} report the scaled Montecarlo results for a target  neutrino mass sensitivity $\Sigma_{90}(m_\nu)$ equal to 0.2\,eV and 0.1\,eV respectively. Results are obtained
in absence of background. The first line is a sort of baseline experimental configuration 
characterized by very demanding energy and time resolution and by very limited pile-up fraction $f_{pp}$ obtained 
by keeping the single detector activity $A_\beta$ at 1\,Hz: in this conditions 
the target sensitivity is achieved with a relatively low statistics $N_{ev}$ at the expenses of a large required exposure $T$.
In the other lines of the tables a larger activity of 10\,Hz is considered together with a progressive degradation of 
energy and time resolution. While a larger activity implies a lower required exposure, the poorer performances are compensated by
the need of a larger statistics. From the tables it is clear that one can find a compromise between performances and exposure which
is more convenient than the baseline high performance experimental configuration. 

For example a target neutrino mass sensitivity of 0.1\,eV could be expected running for 10\,years $3\times10^5$ rhenium
detectors, each with a mass of 10\,mg -- giving an activity of about 10\,Hz --  and with energy and time resolutions of about 1\,eV and 1\,\mus\ respectively. The total required mass of rhenium is about 3\,kg.

\begin{table}[htb!]
\caption{\label{tab:exp1} Experimental exposure required for a 0.2\,eV \mn\ statistical sensitivity.}
\begin{center}
\begin{tabular}{ccccc}
\hline 
$A_\beta$ &	$\tau_R$ &	\de\ &	$N_{ev}$ &	exposure $T$ \\[0pt] %
[Hz] &	[$\mu$s] &	[eV] &	[counts] &	[detector$\times$year] \\
\hline 
1& 1&	1&	$0.2\times10^{14}$&	$7.6\times10^{5}$\\
\hline 
10&	1&	1&	$0.7\times10^{14}$&	$2.1\times10^{5}$\\
10&	3&	3&	$1.3\times10^{14}$&	$4.1\times10^{5}$\\
10&	5&	5&	$1.9\times10^{14}$&	$6.1\times10^{5}$\\
10&	10&	10&	$3.3\times10^{14}$&	$10.5\times10^{5}$\\
\hline 
\end{tabular} 
\end{center}
\end{table}

\begin{table}[htb!]
\caption{\label{tab:exp2} Experimental exposure required for a 0.1\,eV \mn\ statistical sensitivity.}
\begin{center}
\begin{tabular}{ccccc}
\hline 
$A_\beta$ &	$\tau_R$ &	\de\ &	$N_{ev}$ &	exposure $T$ \\[0pt] %
[Hz] &	[$\mu$s] &	[eV] &	[counts] &	[detector$\times$year] \\
\hline 
1& 0.1&	0.1&	$1.7\times10^{14}$&	$5.4\times10^{6}$\\
\hline 
10&	0.1&	0.1&	$5.3\times10^{14}$&	$1.7\times10^{6}$\\
10&	1&	1&	$10.3\times10^{14}$&	$3.3\times10^{6}$\\
10&	3&	3&	$21.4\times10^{14}$&	$6.8\times10^{6}$\\
10&	5&	5&	$43.6\times10^{14}$&	$13.9\times10^{6}$\\
\hline 
\end{tabular} 
\end{center}
\end{table}

\section{Conclusions}
In this paper we have thoroughly discussed the statistical sensitivity of calorimetric Rhenium based neutrino mass experiments.

 To estimate the statistical sensitivity, two methods have been developed. They are based respectively
on an analytic and a Montecarlo approach: the results presented and compared in \S\,\ref{sec:ana-vs-mc} show the prominent importance of the total statistics collected by such an experiment in order to reach a sub-eV sensitivity. 

Extending the application of the Montecarlo approach, we have then analyzed the expected sources of systematic uncertainties peculiar to this kind of experiments.
In particular, in \S\,\ref{sec:source-sys} we have shown how crucial is for future experiments the understanding
of the theoretical \Re\ beta decay spectrum and of BEFS. On the other hand, in \S\,\ref{sec:instr-sys} 
we have shown how instrumental systematic uncertainties may be kept under control by a proper characterization
of the response function $R(E)$ and by an accurate detector calibration: tasks that may be  accomplished by controlling the calibration peak statistic. 

Finally, in \S\,\ref{sec:future}, we have exploited the statistical analysis  to 
devise a plausible experimental configuration capable to achieve a sensitivity of about 0.1\,eV on the neutrino mass.

As a concluding remarks, we believe we have demonstrated that calorimetric neutrino mass experiments with Rhenium based
detectors offer a realistic chance to reach sensitivities comparable, or even beyond, the KATRIN goal. Moreover
we have shown that, although systematics related to  \Re\ beta decay theory and to BEFS require
further investigations, these experiments should not be plagued by large systematic uncertainties.

The authors wish to thank Prof. Dan McCammon for the many stimulating discussions on the topic of this paper.

\end{document}